\documentclass[twocolumn,nofootinbib,amsmath,prd,aps,superscriptaddress,tightenlines,preprintnumbers]{revtex4}

\usepackage{graphicx}
\usepackage{braket}
\usepackage[usenames]{color}
\usepackage{latexsym}
\usepackage{amsmath}
\usepackage{mathrsfs}
\usepackage{hyperref}
\usepackage{enumerate}

\def\bea{\begin{eqnarray}}
\def\eea{\end{eqnarray}}

\preprint{DESY 16-092}

\begin{document}
\newcount\hour \newcount\minute
\hour=\time \divide \hour by 60
\minute=\time
\count99=\hour \multiply \count99 by -60 \advance \minute by \count99
\newcommand{\mydate}{\ \today \ - \number\hour :00}

\title{A unified explanation for dark matter and electroweak baryogenesis\\ with direct detection and gravitational wave signatures}

\author{Mikael Chala${}^a$, Germano Nardini${}^{b}$ and Ivan Sobolev${}^{c, d}$\\\vspace{0.4cm}
\it ${}^a$DESY, Notkestrasse 85, D-22607 Hamburg, Germany\\\vspace{0.1cm}
\it ${}^b$Albert Einstein Center, Institute for Theoretical Physics, University of Bern, Sidlerstrasse 5, CH-3012 Bern, Switzerland\\\vspace{0.1cm}
\it ${}^c$Institute for Nuclear Research of the Russian Academy of Sciences, 60th October Anniversary prospect 7a, Moscow 117312, Russia\\\vspace{0.1cm}
\it ${}^d$Department of Particle Physics and Cosmology, Physics Faculty, M. V. Lomonosov Moscow State University, Vorobjevy Gory, 119991, Moscow, Russia
}

\begin{abstract}
  A minimal extension of the Standard Model that provides both a dark matter candidate and a strong first-order electroweak phase transition (EWPT) consists of two additional Lorentz and gauge singlets. In this paper we work out a composite Higgs version of this scenario, based on the coset $SO(7)/SO(6)$. We show that by embedding the elementary fermions in appropriate representations of $SO(7)$, all dominant interactions are described by only three free effective parameters.
  Within the model dependencies of the embedding, the theory predicts one of the singlets to be stable and responsible for the observed dark matter abundance.
  At the same time, the second singlet introduces new $CP$-violation phases and triggers a strong first-order EWPT, making electroweak baryogenesis feasible. 
  It turns out that this scenario does not conflict with current observations and it is promising for solving the dark matter and baryon asymmetry puzzles. The tight predictions of the model will be accessible at the forthcoming dark matter direct detection and gravitational wave experiments.
\end{abstract}

\maketitle
\newpage

\section{Introduction}

\noindent
In light of the celebrated discovery of the Higgs boson~\cite{Aad:2012tfa,Chatrchyan:2012xdj}, and the absence of further signatures of new physics at the LHC and other experiments (we cannot assert the real nature of the 750 GeV excess~\cite{ATLAS:2015dp,CMS:2015dxe} for the time being), the success of the Standard Model of particle physics (SM) has become unquestionable nowadays. Nonetheless, there are both experimental observations and theoretical reasons not to consider the SM as a complete description of Nature even at low energies. As a matter of fact, the SM does not account for the neutrino oscillations, nor for the evidence of Dark Matter (DM) and the baryon-antibaryon asymmetry of the Universe, among others. On top of that, the unnatural size of the Higgs mass suggests that new degrees of freedom should show up at
energies of the order of the TeV scale. In this spirit, Composite Higgs Models (CHMs)~\cite{Kaplan:1983sm,Kaplan:1983fs,Dimopoulos:1981xc} have played a major role in the recent years. In particular, the minimal CHM~\cite{Agashe:2004rs} has deserved special attention, but non-minimal CHMs have been also studied in detail, as their extended scalar sectors may provide solutions to the intricacies of new physics~\cite{Gripaios:2009pe,Bellazzini:2014yua,Mrazek:2011iu,Espinosa:2011eu,Chala:2012af,Frigerio:2012uc,Bertuzzo:2012ya,Vecchi:2013bja,Fonseca:2015gva,Nevzorov:2015sha}. This is specially well motivated, given that departures from the SM predictions can be plausibly hidden in the Higgs sector, which is not precisely measured yet.

Non-minimal CHMs are very predictive in comparison to their renormalizable counterparts built in a bottom-up approach. 
In CHMs the light scalars are indeed the pseudo-Nambu-Goldstone Bosons (pNGBs) of a spontaneously-broken global symmetry of a new strongly-coupled sector. The derivative and gauge interactions are hence fixed by the coset structure. The leading-order one-loop potential is instead computed as an expansion on spurion insertions parameterizing the explicit breaking of the global symmetry~\cite{Panico:2015jxa}. The crucial point in this regard is that the number of independent spurion invariants is generally smaller than the number of independent interactions compatible with the remnant symmetry, which in addition may be larger than the SM gauge group. A caveat is still in order for that matter, namely small spurion multiplets are typically disfavored. In fact, they use to generate no quartic term for the Higgs boson at leading order; the proper size of such a term arising only by assuming large next-to-leading order corrections, which tend to make the usual perturbative ordering unreliable~\cite{Panico:2015jxa}.

In view of the discussion above, in the present paper we explore the phenomenology of a non-minimal CHM based on the coset $SO(7)/SO(6)$. We show that by taking spurions transforming in the fundamental representation $\mathbf{7}$ and the symmetric representation $\mathbf{27}$ of $SO(7)$, the model provides a rather predictive setup where the low-energy scalar sector consists of a Higgs field and  
two real scalars, $\eta$ and $\kappa$, that are singlets under the SM gauge group. Remarkably, this $SO(7)$ breaking pattern leads to an approximate $Z_2$ symmetry for $\kappa$ and an exact $Z_2$ symmetry for $\eta$. In addition, the field $\kappa$ does not acquire a vacuum expectation value (VEV). It is hence stable and turns out to be a good DM candidate. At the same time, within the (modest) model dependencies of the theory, the singlet $\eta$ acquires a VEV prior to the electroweak symmetry breaking and ultimately promotes a strong first-order electroweak phase transition (EWPT) with sizable gravitational wave signals. Besides, higher-dimensional operators with sizeable  $CP$-violation phases are expected~\cite{Espinosa:2011eu}, opening up the possibility of electroweak baryogenesis \cite{Kuzmin:1985mm} (related results in non-composite models are discussed in refs.~\cite{Cline:2012hg,Cline:2013gha,Profumo:2014opa,Jiang:2015cwa}). Notably, domain wall problems are avoided for the $Z_2$ symmetry of $\kappa$ is not exact.
Hence, within the natural values, the model can potentially solve the hierarchy problem as well as the cosmological evidence of DM and the baryon-antibaryon asymmetry of the Universe. Forthcoming experiments will be able to probe the model predictions.

The rest of this paper is structured as follows. In sec.~\ref{sec:chm} we revisit briefly the basics of CHMs, paying special attention to the computation of the scalar potential. In sec.~\ref{sec:model} we introduce the model. In sec.~\ref{sec:dm} we work out the implications for DM while in sec.~\ref{sec:ew} we study the EWPT.  In sec.~\ref{sec:constraints} we discuss the current constraints and future probes, including gravitational wave searches, Higgs physics, DM direct-detection experiments, and LHC searches for DM and dijet resonances. Sec.~\ref{sec:conclusions} is devoted to conclusions.

\section{A quick look at composite models}
\label{sec:chm}

\noindent
In this section we present some introductory ideas on CHMs and set conventions up. Readers interested in further details are referred to e.g.~the recent review \cite{Panico:2015jxa}, whereas those familiar with CHMs may go directly to sec.~\ref{sec:model}. 

CHMs extend the SM with a new strongly-interacting sector characterized by a typically scale $f\sim\rm{TeV}$ and a coupling $g$. This new sector is endowed with an approximate global symmetry group, $G$, spontaneously broken into some subgroup, $H$, containing the SM gauge group. 
The Higgs boson is assumed to be a pNGB of this symmetry-breaking pattern, and hence naturally lighter than $f$. 

The global symmetry is assumed to be explicitly broken mainly by the linear mixings between the SM fermions and composite operators $\mathcal{O}$ of the strong sector. The dynamics of the Higgs and other potential pNGBs is dictated by the following Lagrangian:
\begin{equation}
 \mathcal{L} = \mathcal{L}_\sigma + \mathcal{L}_{y} - V~.
\end{equation}
$\mathcal{L}_\sigma$ describes the gauge and (derivative) scalar interactions. It is fixed by the coset structure $G/H$. At the leading order in the derivative expansion, it reads
\begin{equation}
 \mathcal{L}_\sigma = \frac{f^2}{4} \text{Tr} (d_{\mu} d^{\mu})~,
\end{equation}
where $d_\mu$ is the projection of the Maurer-Cartan one-form 
 $\omega_\mu = i U^{-1} D_\mu U$
into the broken generators $T^i$, with $U$ being the Goldstone matrix
\begin{equation}
 U[\Pi] = \exp \left \{ -i \frac{\sqrt{2}}{f} \, h_i \, T^i \right \}
\end{equation}
and $h_i$ the pNGBs. $\mathcal{L}_y$ and $V$ stand for the Yukawa Lagrangian and the scalar potential, respectively. They arise from the explicit breaking of the global symmetry. According to the partial compositeness setup~\cite{Kaplan:1991dc}, such a breaking is provided by
\begin{equation}
\mathcal{L}_\text{mix} = \sum_{\Psi} \lambda_{\Psi}\overline{\Psi} \mathcal{O}_\Psi +\text{h.c.}~,
\end{equation}
with $\Psi$ running over incomplete representations of $G$ (also called \textit{spurions}) embedding the SM fermions. 

The so-called dressed fields $\Psi^D$ can be constructed out of the spurions as	
\begin{equation}
 \Psi^D \equiv U^{-1} \Psi~.
\end{equation}
In general, $\Psi^D$ transform in reducible representations of the unbroken group $H$, namely
\begin{equation}
 \Psi^D = \bigoplus\Psi^D_m~,
\end{equation}
with $m$ running over the irreducible representations of $H$.

The dressed fields are useful to determine the invariants of $H$. By calling $I_n^j$ the $j$-th invariant of $H$ involving a number $n$ of fields, the scalar potential can be expressed as the following expansion~\cite{Panico:2015jxa}: 
\begin{equation}
 V \sim \left(\frac{\lambda_\Psi}{g}\right)^2 \sum_j c_2^j I_2^j + \left(\frac{\lambda_\Psi}{g}\right)^4 \sum_j c_4^j I_4^j + \cdots~,
\end{equation}
where $c_m^j$ are in practice free parameters.  
Given that $\lambda_\Psi$ is expected to be much smaller than the new strong coupling $g$, $I_2^j$ dominate the previous expansion. However, when $\Psi$ transforms in a small representation of $G$, the leading order invariants $I_2^j$ do not usually generate the Higgs quartic coupling, and hence  electroweak symmetry breaking (EWSB) can not be achieved at the leading order~\footnote{For instance,  in the minimal CHM this occurs for the $\mathbf{4}$ and $\mathbf{5}$ representations but not for the $\mathbf{14}$ one.}. 
Small representations can be still considered if one assumes that $V$ is not dominated by the (formally) leading-order contribution. The problem of this regime is twofold. On one hand, the fine-tuning for keeping the leading-order contribution small is obviously large. On the other hand, sizeable next-to-leading order contributions come at the expense of predictivity. Indeed, much less spurions, and hence free parameters, are present at the leading order. For the sake of example, two versus fifteen  independent spurions arise at the leading and next-to-leading order, respectively, in the $\mathbf{5}$ of $SO(5)$~\cite{Panico:2015jxa}.

In the present work  we proceed in the regime where the next-to-leading contributions are subleading and we hence neglect them unless otherwise stated. As previously discussed, we work out the coset $SO(7)/SO(6)$.  We embed the SM fermions in the $\mathbf{7}$ and $\mathbf{27}$ of $SO(7)$. The latter arises as the symmetric part of $\mathbf{7}\times\mathbf{7} = 1 + \mathbf{21} + \mathbf{27}$, in complete analogy with the $\mathbf{14}$ in $SO(5)$. Under $SO(6)$, we obtain the following branching rules:
\begin{equation}
 \mathbf{7} = 1 + \mathbf{6}~,\quad
 \end{equation}
 \begin{equation}
 \mathbf{27} = 1 + \mathbf{6} + \mathbf{20}~.
\end{equation}
One and two independent spurion invariants can  be therefore constructed at the leading order from the $\mathbf{7}$ and $\mathbf{27}$ representations, respectively.

\section{Model description}
\label{sec:model}

\noindent
The model we analyze is based on the symmetry-breaking pattern $SO(7)\times U(1)^\prime/SO(6)\times U(1)^\prime$. We proceed in the unitary gauge. Two gauge singlets, $\eta$ and $\kappa$, arise in the pNGB spectrum in addition to the Higgs degrees of freedom $\phi = [\phi^+, (h+i\phi^0)/\sqrt{2}]^\text{T}$. The addition of a spectator group $U(1)^\prime$ is required in order the SM-fermion hypercharges to be correctly reproduced, in the same vein as in the minimal CHM. 

The fifteen unbroken and six broken generators of $SO(7)$, $T$ and $X$ respectively, can be conveniently written as
\begin{eqnarray}
 T^{mn}_{ij} &=& -\frac{i}{\sqrt{2}} (\delta^m_i\delta^n_j-\delta^n_i\delta^m_j)~, ~~ m<n\in[1,7]~, \nonumber\\
  X^{m7}_{ij} &=& -\frac{i}{\sqrt{2}} (\delta^m_i\delta^7_j-\delta^7_i\delta^m_j)~, ~~\, m\in[1, 6]~.
\end{eqnarray}
The SM $SU(2)_L\times U(1)_Y$ gauge group is thus generated by
\begin{eqnarray}
 J_L^1 = \frac{1}{\sqrt{2}}(T^{14}+T^{23})~, \quad J_L^2 = \frac{1}{\sqrt{2}}(T^{24}-T^{13})~, \nonumber\\
  J_L^3 = \frac{1}{\sqrt{2}}(T^{12}+T^{34})~, \quad J_R^3 = \frac{1}{\sqrt{2}}(T^{12}-T^{34})~,
\end{eqnarray}
being the hypercharge defined as $Y = J_R^3 + Y^\prime$ with $Y^\prime$ the generator of $U(1)^\prime$.

The dynamics of the pNGBs is described by the Goldstone matrix
\begin{equation}
 U = \exp{\left\lbrace-i\frac{\sqrt{2}}{f}\bigg[ T^{47}h + T^{57}\eta + T^{67}\kappa\bigg]\right\rbrace}~.
\end{equation}
After performing the replacements~\cite{Gripaios:2009pe}
\begin{align}
\frac{h^2}{h^2+\eta^2+\kappa^2}\sin^2{\left\lbrace\left[\frac{h^2+\eta^2+\kappa^2}{f^2}\right]^{-\frac{1}{2}}\right\rbrace}\rightarrow h^2,\\
  \frac{\eta^2}{h^2+\eta^2+\kappa^2}\sin^2{\left\lbrace\left[\frac{h^2+\eta^2+\kappa^2}{f^2}\right]^{-\frac{1}{2}}\right\rbrace}\rightarrow \eta^2,\\
 \frac{\kappa^2}{h^2+\eta^2+\kappa^2}\sin^2{\left\lbrace\left[\frac{h^2+\eta^2+\kappa^2}{f^2}\right]^{-\frac{1}{2}}\right\rbrace}\rightarrow \kappa^2,\\\nonumber
\end{align}
we bring $U$ to the two-block matrix form
\begin{equation}
 U = \left(\begin{array}{ccccccc}
            \mathbf{1}_{3\times 3} & & & & & & \\
            & & & 1-\frac{h^2}{1+\Sigma} & -\frac{h\eta}{1+\Sigma} & -\frac{h\kappa}{1+\Sigma} & h\\
            & & & -\frac{h\eta}{1+\Sigma} & 1-\frac{\eta^2}{1+\Sigma} & -\frac{\eta\kappa}{1+\Sigma} & \eta\\
            & & & -\frac{h\kappa}{1+\Sigma} & -\frac{\eta\kappa}{1+\Sigma} & -\frac{\kappa^2}{1+\Sigma} & \kappa\\
            & & & -h & -\eta & -\kappa & \Sigma
           \end{array}\right)~,
\end{equation}
where
\begin{equation}
 \Sigma = \sqrt{1-h^2-\eta^2-\kappa^2}~.
\end{equation}
The sigma model interactions are thus described by the Lagrangian
\begin{equation}\label{eq:sigma}
 \mathcal{L}_{\sigma} = \mathcal{L}_\text{kinetic} + \frac{1}{2}\frac{\left(h\partial_\mu h + \eta\partial_\mu \eta + \kappa\partial_\mu \kappa\right)^2}{f^2-h^2-\eta^2-\kappa^2}~.
\end{equation}
At the level of the sigma model there is a $Z_2\times Z_2\times Z_2$ symmetry given by $h\rightarrow -h$, $\kappa\rightarrow -\kappa$ and $\eta\rightarrow -\eta$, for the coset is symmetric. This symmetry will be only broken by the external sources, which will also induce a potential for the pNGBs.

At the renormalizable level, the most general potential for $\eta$ (stable), $\kappa$ and the Higgs boson $h$ reads
\begin{align}
 V = &-\frac{1}{2}\mu_h^2 h^2 + \frac{1}{2}\mu_\eta^2\eta^2 + \frac{1}{2}\mu_\kappa^2\kappa^2 \nonumber \\
& + \frac{1}{3}A_{\kappa h} \kappa h^2 + \frac{1}{3}A_{\kappa\eta}\kappa\eta^2 + \frac{1}{3}A_\kappa\kappa^3 \nonumber\\ 
 &+ \frac{1}{4}\lambda_h h^4 + \frac{1}{4}\lambda_\eta\eta^4 + \frac{1}{4}\lambda_\kappa \kappa^4 \nonumber\\
 &+\frac{1}{4}\lambda_{h\eta}h^2\eta^2 + \frac{1}{4}\lambda_{h\kappa}h^2\kappa^2 + \frac{1}{4}\lambda_{\eta\kappa}\eta^2\kappa^2~,
\end{align}
which involves 12 independent parameters. Not all of them will be however generated in the present composite setup, at least at the (unsuppressed) leading order. In particular, if we want $\kappa$ to lead to a two-step EWPT and $\eta$ to be a DM candidate without conflicting with Higgs searches (see secs.~\ref{sec:dm} and~\ref{sec:ew}), the following conditions must hold: \textit{(i)}  $\eta\rightarrow -\eta$ is an unbroken symmetry;  \textit{(ii}) $\mu^2_\kappa < 0$; and \textit{(iii}) the physical masses of  $h$ and $\kappa$ are such that $m_h<2 m_\kappa$, which is favored by  $\lambda_{h\kappa} \gtrsim \lambda_h$.

In the present composite scenario, a minimal content satisfying the three above conditions consists of the mixing Lagrangian 
\begin{align}\nonumber
 \mathcal{L}_{\rm mix} = \sum_{\Psi} \lambda_{\Psi_R}\overline{\Psi_R}^I \left(\mathcal{O}^R_\Psi\right)_I + \sum_{\Psi^\prime} \lambda_{\Psi_L}\overline{\Psi_L}^I \left(\mathcal{O}^L_\Psi\right)_I +\text{h.c.}~,
\end{align}
where the first sum extends over $\Psi = T,B,C$ and the second over $\Psi^\prime =Q^t,Q^b,Q^c$, with $T_R$ and $C_R$ transforming in complete singlets of $SO(7)$ with $U(1)^\prime$ charge $2/3$, $B_R$ and $Q_L^b$ transforming in fundamental representations $\mathbf{7}$ of $SO(7)$ with $U(1)^\prime$ charges $-1/3$, and $Q_L^t$ and $Q_L^c$ transforming instead in the symmetric representation $\mathbf{27}$ of $SO(7)$ that results from the tensor product of fundamental representations, i.e. $\mathbf{7}\times\mathbf{7} = 1 + \mathbf{21} + \mathbf{27}$. The most general embedding fulfilling these assignments is explicitly provided by~\footnote{See refs.~\cite{Gripaios:2009pe,Mrazek:2011iu,Frigerio:2012uc, Chala:2012af} for $Z_2$-preserving embeddings in other models of composite DM.} 
\begin{equation}
B_R = \left(\begin{array}{@{}lllllll@{}}
                    0 & 0 & 0 & 0 & 0 & i\gamma b_R & b_R \\
                  \end{array}\right)^\text{T},\\
\end{equation}
\begin{equation}
 Q^b_L = \frac{1}{\sqrt{2}}\left(\begin{array}{@{}lllllll@{}}
                    -it_L & t_L & ib_L & b_L & 0 & 0 & 0 \\
                  \end{array}\right)^\text{T}, 
\end{equation}
\begin{equation}
  Q_L^t = \dfrac{1}{2}\left(\begin{array}{cccccccccc}
            \mathbf{0}_{6\times 6} & & & & & & it_L \\
             & & & & & &t_L \\
             & & & & & &ib_L \\
             & & & & & &-b_L \\
             & & & & & &0 \\
             & & & & & &0 \\
             it_L & t_L & ib_L & -b_L & 0 & 0 & 0 \\
           \end{array}\right) ~, \\[5mm]
\end{equation}
\begin{equation*}
  Q^c_L = \frac{1}{2}\left(\begin{array}{ccccccccc}
            \mathbf{0}_{5\times 5} & & & & & \zeta c_L & ic_L \\
             & & & & &-i \zeta  c_L  & c_L \\
             & & & & & \zeta  s_L & is_L \\
             & & & & & i\zeta s_L & -s_L \\
             & & &  & &        0 & 0\\
             \zeta c_L & -i\zeta c_L & \zeta s_L & i\zeta s_L & 0 & 0 & 0 \\
             ic_L & c_L & is_L & -s_L & 0 & 0 & 0 \\
           \end{array}\right) ~.
\end{equation*}
Heavier quarks couple stronger to the composite sector, and hence $b$ and $c$ contributions to the one-loop potential can be neglected unless they are multiplied by a large $\gamma$ or $\zeta$, respectively. 

In this embedding the potential acquires the form
\begin{align}\label{eq:potential}
  V =& -\frac{1}{2}\mu_h^2 h^2  + \frac{1}{2}\mu_\eta^2\eta^2 + \frac{1}{2}\mu_\kappa^2\kappa^2 \nonumber\\
 &+ \frac{1}{4}\lambda_h h^4 + \frac{1}{4}\lambda_\kappa\kappa^4 + \frac{1}{4}\lambda_{h\eta} h^2\eta^2 + \frac{1}{4}\lambda_{h\kappa}h^2\kappa^2~.
\end{align}
The quartic coupling $\lambda_\kappa$ is generated only at the next-to-leading order, but it has been introduced since it plays an important role in the EWPT phenomenology. At any rate, it is expected to be much smaller than the other quartic couplings. The rest of the parameters are functions of the dimensionless spurion coefficients $\alpha_{q,i}$, as well as $\gamma$ and $\zeta$:
\begin{align}
 \mu_h^2 &= -\frac{1}{2}f^2\left(4\alpha_{t,1} - 7\alpha_{t,2} + \alpha_{c,2}\zeta^2\right)~,\\
  \mu_\eta^2 &= -2\alpha_{t,2} f^2~,\\
  \mu_\kappa^2 &= 2f^2\left(\alpha_b\gamma^2 + \alpha_{c,2}\zeta^2-\alpha_{t,2}\right)~,\\
  \lambda_h &=4(\alpha_{t,2}-\alpha_{t,1})~,\\
 \lambda_{h\eta} & = 4(\alpha_{t,2}-\alpha_{t,1})~,\\
 \lambda_{h\kappa} &= 4\left[\alpha_{t,2} - \alpha_{t,1} + (\alpha_{c,1}-\alpha_{c,2})\zeta^2\right]~.
\end{align}

In our analysis we will consider two broad parameter regimes depending on the actual values of $\alpha_{c,i}$:
\begin{description}
  \item[\textbf{Regime I}:] $\alpha_{c,2} = -\alpha_{c,1}$. This is the most natural scenario since the size of these two coefficients is expected to be similar, and it still allows for $\lambda_{h\kappa} \gtrsim \lambda_h$, contrary to the case $\alpha_{c,2} = \alpha_{c,1}$.  
  \item[\textbf{Regime II}:] $|\alpha_{c,2}| \ll |\alpha_{c,1}|\sim |\alpha_{t,i}/\zeta^2|$.  As we will see, accounting for the DM relic density observation will completely fix the mass of $\eta$ and its interactions with nuclei in this case~\footnote{The case $|\alpha_{c,1}| \ll |\alpha_{c,2}|$ would be quite similar to \textit{Regime I}.
}.
\end{description}
In both cases, the coefficients $\alpha_q^i$ as well as $\gamma$ and $\zeta$ can be traded by the measured values of the Higgs VEV ($v\simeq 246$ GeV) and physical mass (yielding $\lambda\simeq 0.13$), and only three free parameters, namely $f$, $\mu_\kappa^2$ and $\lambda_{h\kappa}$. Indeed, in \textit{Regime I} we obtain
\begin{align}\label{eq:reg1}
 \mu_\eta^2 &= \frac{1}{3}f^2\left[\frac{7}{4}\lambda_h+\frac{1}{4}\lambda_{h\kappa}-4\lambda_h\xi\right]~,\\
 \lambda_{h\eta} &= \lambda_h~,
\end{align}
while in \textit{Regime II} we get
\begin{align}\label{eq:reg2}
\mu_\eta^2 &= \frac{2}{3}\lambda_h f^2\left(1-2\xi\right)~,\\
\lambda_{h\eta} &= \lambda_h~,
\end{align}
with $\xi\equiv v^2/f^2$.

Finally, the Yukawa Lagrangian takes the form
\begin{align}\label{eq:yukawa}\nonumber
 \mathcal{L}_y =& -\sum_{q = t,b,c}y_q\overline{q}qh\left[1-\frac{1}{f^2}\left(h^2+\eta^2+\kappa^2\right)\right]^\frac{1}{2}\\
 &-i \frac{h}{f}\kappa\left[\gamma y_b\overline{b}\gamma_5 b +  \zeta y_c\overline{c}\gamma_5 c\right]~,
\end{align}
with $y_q$ the Yukawa couplings.

According to these results, several comments are in order here:
\begin{enumerate}[i)]
\item Neither a single cubic coupling nor a quartic term for $\eta$ or a quartic coupling between $\eta$ and $\kappa$ are generated at leading order. At higher orders only the $Z_2$ symmetry of $\kappa$ is broken. Therefore, since $\mu_\eta^2$ is predicted to be positive, $\eta$ is stable and hence a DM candidate. Finally, $\kappa$ is subject to model dependencies that allow $\mu^2_\kappa<0$~\footnote{Relevant comments in this respect have been previously pointed out in ref.~\cite{Serra:2015xfa}.}.
\item The second quark generation has to be included  in order the coupling $\lambda_{h\kappa}$ to be large enough to achieve the hierarchy $m_h^2/4 < m^2_\kappa = \mu^2_\kappa + 1/2\lambda_{h\kappa} v^2$ between the  $h$ and $\kappa$ physical masses. Otherwise the experimental bound on the Higgs decay into non-SM particles~\cite{Atlas:2015at}  would be hard to evade.
\item $\gamma$ and $\zeta$ could have a small imaginary part, even in the top sector,
without susbtantially changing the equations above. If this was the case, they
could provide a  new sizable source of CP violation~\cite{Espinosa:2011eu}, as required by electroweak baryogenesis.
\item From the EWSB conditions one obtains $|\alpha_{t,i}|\sim \lambda_h\simeq 0.13$\,. We besides expect $|\alpha_c| < |\alpha_b| <|\alpha_{t,i}|$. We have checked that, in both regimes, independently of $f \le 5$ TeV, any value of $m_\kappa$ below 200 GeV and any value of $\lambda_{h\kappa}$ between 0.1 and 0.4 can be reached by barring this parameter space region with mild values of $\gamma,\zeta\in [3, 5]$.
\end{enumerate}

\section{Dark matter predictions}
\label{sec:dm}

\begin{figure}[t]
 \includegraphics[width=\columnwidth]{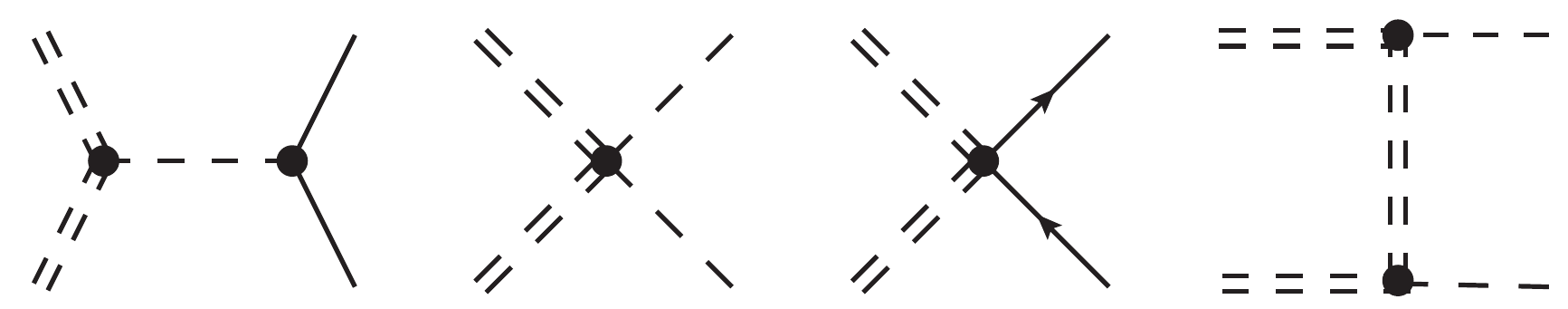}
 \caption{Main diagrams contributing to the DM annihilation. The double dashed lines stand for $\eta$. In the first plot the solid legs account for either SM particles or $\kappa$. In the second diagram the simple dashed lines represent either the Higgs boson or $\kappa$. In the third diagram the fermion lines stand for (mainly) the top quark. In the fourth diagram the simple dashed lines represent the Higgs boson.}\label{fig:dmdiag}
\end{figure}

\noindent
As previously discussed, $\eta$ can provide a DM candidate since it is protected by a $Z_2$ symmetry not even spontaneously broken.  The main diagrams contributing to the annihilation of $\eta$ are shown in fig.~\ref{fig:dmdiag}.  For the choice $f\sim$~TeV that is favored by electroweak precision data and Higgs physics~(see sec.~\ref{sec:constraints}),  all depicted processes are kinematically accessible. Indeed, as eqs.~(\ref{eq:reg1}) and (\ref{eq:reg2}) show, the physical mass of $\eta$, $m_\eta$, is larger than the electroweak scale (and thus than the quark and remaining pNGB masses) when $f\gg\,v$. Due to this hierarchy, the DM phenomenology is dominated by one single scale, $m_\eta$ (or equivalently $f$). No strong dependence on $m_\kappa$ is thus expected. 

One can use dimensional analysis to estimate the annihilation cross sections of $\eta$. As the processes with mediators are suppressed, from eqs.~(\ref{eq:sigma}), (\ref{eq:potential}) and (\ref{eq:yukawa}) one deduces
\begin{align}
\label{eq:ann_hh}
  \sigma(\eta\eta\rightarrow h h) v_0  &\sim \frac{1}{m_\eta^2}\left[\lambda_h - \frac{4m_\eta^2}{f^2}\right]^2,\\
\label{eq:ann_kk}
 \sigma(\eta\eta\rightarrow \kappa\kappa) v_0 &\sim \frac{1}{m_\eta^2}\left[\frac{4m_\eta^2}{f^2}\right]^2,\\
\label{eq:ann_tt}
 \sigma(\eta\eta\rightarrow t\bar{t}) v_0 &\sim \frac{1}{m_\eta^2}\left[\frac{m_t m_\eta}{f^2}\right]^2,
\end{align}
whit $v_0$ the (small) velocity of the colliding DM particles.
It is then expected that the $\eta\eta\rightarrow \kappa\kappa$ channel dominates the annihilation cross section,
 for a partial cancellation between the Higgs quadratic coupling and the derivative contribution from $\mathcal L_\sigma$ arises in $ \sigma(\eta\eta\rightarrow h h)$.  This implies that the DM phenomenology can be drastically different from related composite setups.

\begin{figure}[t]
  \hspace{-0.5cm}
  \includegraphics[scale=0.66]{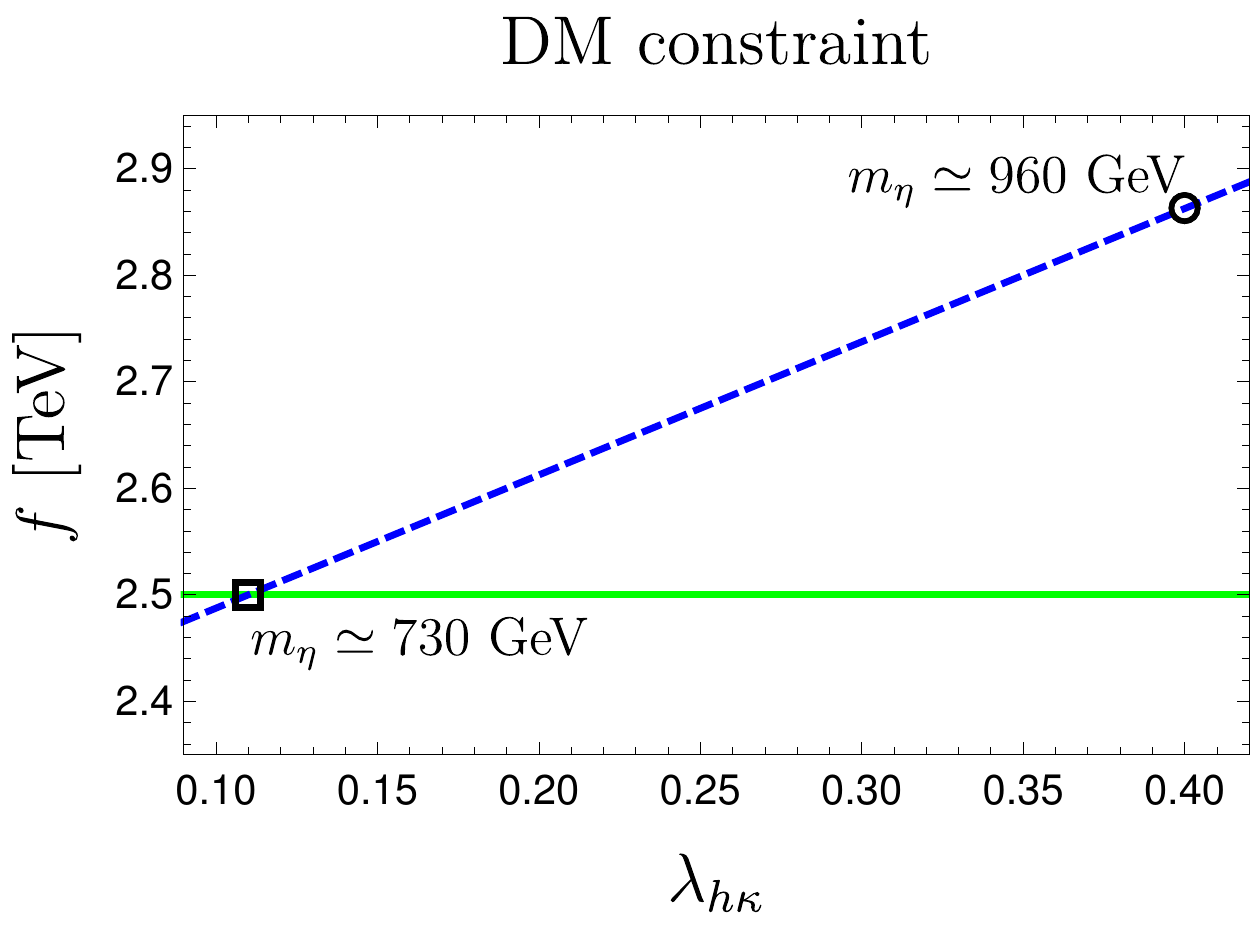}
 \caption{Value of $f$ leading to $\Omega_\eta  = \Omega_{\rm DM}$ as a function of $\lambda_{h\kappa}$ in \textit{Regime I} (dashed blue) and \textit{Regime II} (solid green). The masses $m_\eta$ corresponding to two extreme points are also depicted.}\label{fig:dm}
\end{figure}

Interestingly, in our framework the DM abundance is fixed by only a few free parameters. Indeed, since $\sigma(\eta\eta\rightarrow \kappa\kappa)$ is the dominant 
annihilation cross section, the relic density depends only on $\lambda_{h\kappa}$ and $f$ in \textit{Regime II}, and uniquely on $f$ in \textit{Regime II} (cf.~eqs.~\eqref{eq:reg1} and \eqref{eq:reg2}).  This feature also arises in our numerical study. Specifically, we employ  \texttt{Feynrules}~\cite{Alloul:2013bka} to implement the model, and \texttt{micrOmegas}~\cite{Belanger:2014vza} to determine the parameter region where the relic abundance of $\eta$, $\Omega_\eta h^2$, is compatible with the experimental measurement  $\Omega_\text{DM} h^2 = 0.119\pm 0.003$~\cite{Ade:2013zuv}. The finding is summarized in fig.~\ref{fig:dm} which highlights the constraint $f[\lambda_{h\kappa}]$ that guarantees $\Omega_\eta h^2 = 0.119$~ in \textit{Regime I} (dashed blue) and \textit{Regime II} (solid green). Clearly, for $\lambda_{h\kappa} = \lambda_h$ the expected value of $f$ is the same in both regimes. The figure also displays two successful parameter points with somehow extreme values of $\lambda_{h\kappa}$, and the corresponding predictions for $m_\eta$. In conclusion, due to the DM relic density constraints, we expect $m_\eta\simeq 730\div 960$ GeV and $f\simeq 2.4\div 2.9$ TeV. 

\section{Electroweak baryogenesis}
\label{sec:ew}
\noindent
As generically expected in composite models, and commented above, the interactions between  the quarks and the strong sector can contain new CP-violation phases. It was observed that these phases, together with the sphaleron processes,  can lead to the observed baryon asymmetry,  provided the EWPT is of strongly first order~\cite{Espinosa:2011eu}.  Here we demonstrate that, in our composite model, such a strong EWPT is rather likely within the parameter region compatible with the present constraints, DM bounds included~\footnote{A phase transition, possibly linked to the baryon asymmetry production~\cite{Nardini:2007me, Konstandin:2011ds}, may also occur during the breaking of the composite strong symmetry~\cite{Creminelli:2001th, Randall:2006py, Konstandin:2010cd, Konstandin:2011dr, Hassanain:2007js, Schwaller:2015tja}. Such a transition would modify our results only if, in the present setup, it turned out to be of first order, very supercooled, and with a reheating temperature around or below the electroweak scale, characteristics that may or may not be realized depending on the ultraviolet completion~\cite{Creminelli:2001th,Hassanain:2007js} and its parameter values~\cite{Nardini:2007me,Konstandin:2011ds}.}.

We deduce the properties of the EWPT from the one-loop finite-temperature scalar potential  $V_{\rm{1L}}(h,\kappa,\eta\,; T)$, with $T$ representing the temperature. Since the new fields have small couplings and negligible mixing with the Higgs, at qualitative level $V_{\rm{1L}}(h,0,0; T\!=\!0)$ is similar to the potential in eq.~\eqref{eq:potential}. Therefore, in the sizeable part of the fundamental parameter space predicting $\mu^2_\kappa<0$ and $\lambda_{h\kappa}>0$, the minima of $V_{\rm{1L}}(h,\eta,\kappa; T\!=\!0)$ are
\begin{eqnarray}
 &&v_1(T=0) = (v[T\!=\!0],0,0)~,\\  
&&v_2(T=0)= (0,v_\kappa[T\!=\!0],0)~, \label{eq:v1v2}
\end{eqnarray}
with $v_\kappa[T\!=\!0] \simeq \sqrt{-\mu^2_\kappa/\lambda_\kappa}$ and $v[T\!=\!0] = v$~\footnote{It is not restrictive to ignore the other symmetric minima. Moreover, due to the (suppressed) explicit breaking of the $\kappa$ discrete symmetry, only one of the two minima $\pm v_2$ will be relevant in the evolution of the Universe~\cite{Espinosa:2011eu}.}. 

This structure of the minima hints at the possibility of a two-step EWPT. In this case the electroweak breaking minimum is reached via the changes of phases $(0,0,0)\to v_2(T^\prime)$ and $v_2(T_n)\to v_1(T_n)$,  with $T^\prime>T_n$. The latter transition is the one that is required to be strong (i.e.~$|v_1(T_n)|/T_n>1$) for successful electroweak baryogenesis. 

To determine the possible phase transitions and their characteristics, we use \texttt{CosmoTransitions}~\cite{Wainwright:2011kj}. In the code we specify $V_{\rm{1L}}(h,\kappa,\eta\,; T)$ in the customary form~\cite{Quiros:1994dr}
\begin{eqnarray}
  V_{\rm{1L}}(h,\kappa; T)=V + \Delta V_{\rm{CW}} + \Delta V_{T\neq0}~,
\end{eqnarray}
with $V$ given in eq.~\eqref{eq:potential} and 
\begin{eqnarray}
  \Delta V_{\rm{CW}} &=& \frac{1}{64\pi^2}\sum_i (\pm1) n_i m_i^4(h,\kappa)\bigg[\log{\frac{m_i^2(h,\kappa)}{v^2}-c_i}\bigg]~,\nonumber \\
  \Delta V_{T\neq 0} &=& \frac{T^4}{2\pi^2}\sum_i (\pm1) n_i~ J_\pm\!\left(\frac{m_i^2(h,\kappa)}{T^2}\right)~, \nonumber \\
J_\pm(x) &=& \int_0^\infty dy y² \left[ 1 \mp \exp \left( -\sqrt{x^2+y^2}  \right) \right]~,
\label{eq:deltaV}
\end{eqnarray}
in which the dependence on the background field of $\eta$ is removed because $\partial^2_\eta  V>0$ for any VEV of $h$ and $\kappa$ below $f$. 
In eq.~\eqref{eq:deltaV} $i$ extends to the fields that couple stronger to the Higgs sector, namely the $\kappa$-$h$ mixing states $\phi_{1,2}$, the SM Goldstones $G^{0,\pm}$, the singlet $\eta$, the top quark $t$, and the gauge bosons $W^\pm$ and $Z$. The factor $n_i$ is the number of degrees of freedom of the field $i$, and the upper sign (lower sign) in `$\pm$' and `$\mp$' applies to the bosonic (fermionic) contributions.  The coefficient $c_i$ is equal to 5/6 for gauge bosons and to 3/2 otherwise.  Finally, the field-dependent squared masses $m_{\phi_1}^2(h,\kappa)$ and  $m_{\phi_1}^2(h,\kappa)$ are the eigenvalues of the symmetric matrix $\mathcal M^2$ with entries 
\begin{eqnarray}
\mathcal M^2_{1,1} &=&   -\mu_h^2 + 3\lambda_h h^2 +\frac{1}{2}\lambda_{h\kappa} \kappa^2~,\\
\mathcal M^2_{2,2} &=&   \mu_\kappa^2 + 3\lambda_\kappa \kappa^2 +\frac{1}{2}\lambda_{h\kappa} h^2~,\\
\mathcal M^2_{1,2} &=&   \lambda_{h\kappa} h \kappa~,
\end{eqnarray}
while the other masses are
\begin{align}
 m_{G^{0,\pm}}^2(h,\kappa) &= -\mu_h^2 + \lambda_h h^2 +\frac{1}{2}\lambda_{h\kappa} \kappa^2~,\\
 m_W^2(h,\kappa) &= \frac{1}{4}g^2h^2~,\\
 m_Z^2(h,\kappa) &= \frac{1}{4}(g^2 + g'^2)h^2~,\\
 m_t^2(h,\kappa) &= \frac{1}{2}y_t^2h^2~,\\
m_\eta^2(h,\kappa) &= C(h)~,
 \end{align}
where $g, g'$ and $y_t$ stand for the $SU(2)_L$ and $U(1)_Y$ gauge couplings and the top Yukawa, respectively. The function $C(h)$ depends on which regime we consider. Taking into account the DM constraint that establishes the function $f[\lambda_{h\kappa}]$ (see fig.~\ref{fig:dm}), we have
 \begin{equation}
C(h) = \frac{1}{3}\left[\frac{7}{4}\lambda_h+\frac{1}{4}\lambda_{h\kappa}-4\lambda_h\xi\right]  f^2[\lambda_{h\kappa}]+ \frac{1}{2}\lambda_h h^2 \\
 \end{equation}
in \textit{Regime I} and 
\begin{equation}
 C(h) = \frac{1}{2}\lambda_h\left[\frac{4}{3}(1-2\xi) f^2[\lambda_{h\kappa}]+ h^2\right]
\end{equation}
in \textit{Regime II}.
\begin{figure*}[t]
  \centering
\hspace{-10mm}\includegraphics[width=\columnwidth]{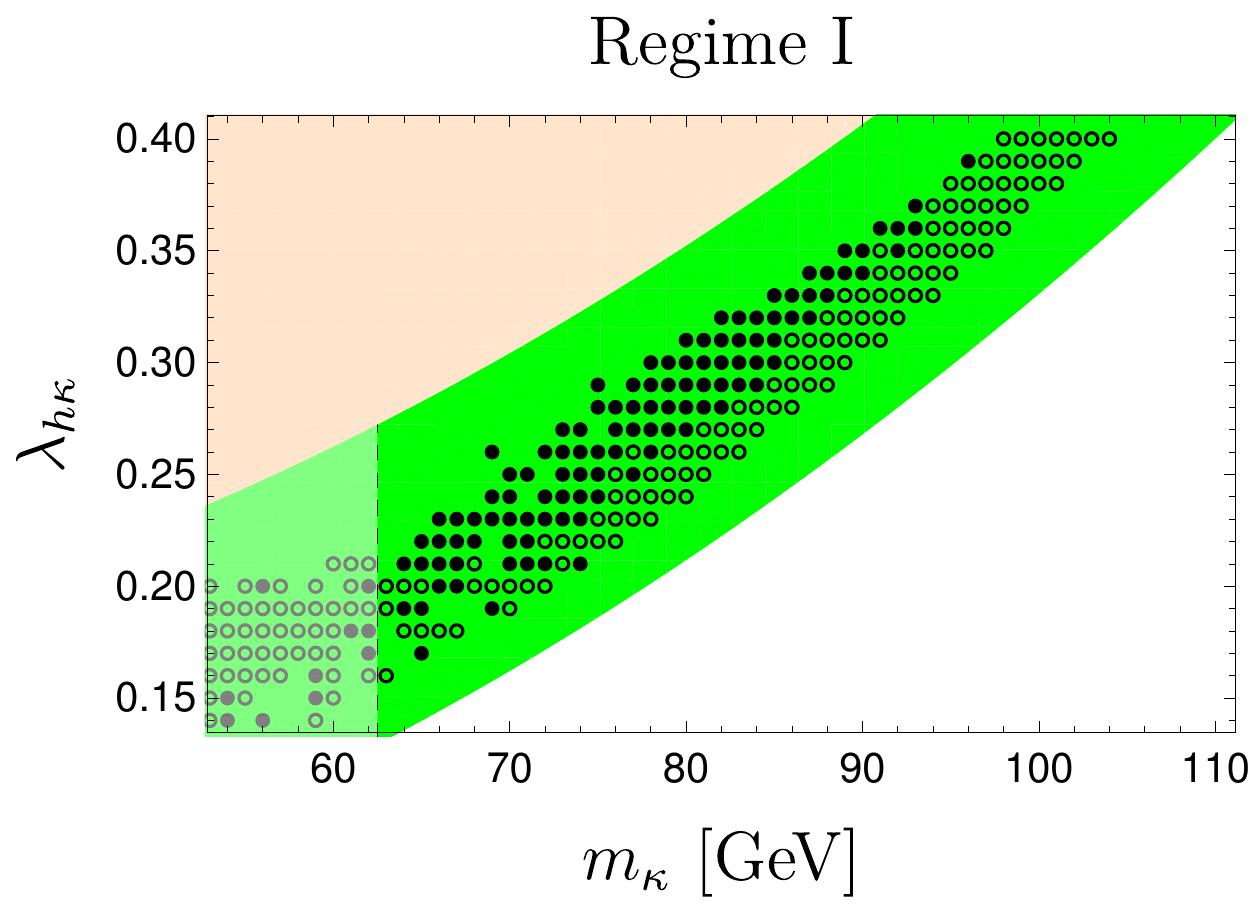}
\includegraphics[width=\columnwidth]{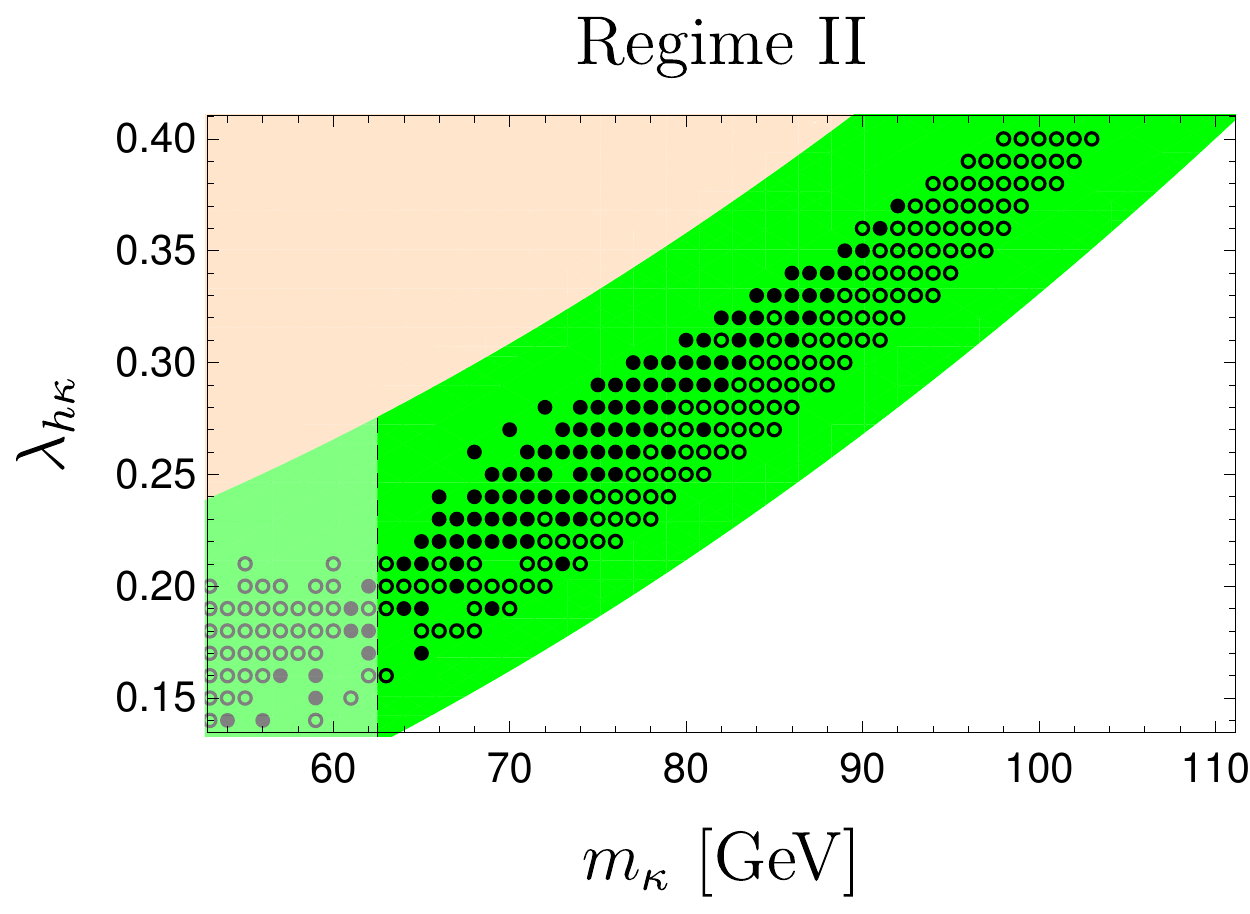}
\caption{Points of the parameter space exhibiting a strong EWPT in \textit{Regime I} (left panel) and \textit{Regime II} (right panel). The region in green indicates the points for which $V_{\rm 1L}(h,\kappa; T\!=\!0)$ has  a local minimum at $v_1$ (the contrary in the white area) and such a minimum is deeper than the one at $v_2$ (the contrary in the orange area).   The points on the left of the black dashed line are unfavored by the Higgs searches. The filled (empty) circles correspond to EWPTs with bubbles expanding (not expanding) at the speed of light.}\label{fig:ewpt}
\end{figure*}

For the numerical analysis we scan over the parameter region $\sqrt{-\mu^2_\kappa}/{\rm GeV}\in\, ]0, 100]$ and $\lambda_{h\kappa}\in[0.1, 0.4]$. For each pair $\{\mu_\kappa^2,\lambda_{h\kappa}\}$ we consider the lowest value of $\lambda_{\kappa}\in[0.01, 0.02, \dots, 0.06]$ for which the EWPT is strong, if such a transition arises~\footnote{Due to some numerical instabilities, the code identifies only a subset of the satisfactory parameter points. Since our aim is to highlight the abundance of points with a strong first-order EWPT, we do try to circumvent this issue and we just display the numerous points that the code finds.}. The findings are presented in fig.~\ref{fig:ewpt} for  \textit{Regime I} (left panel) and \textit{Regime II} (right panel). Both filled and empty circles represent the parameter points where the EWPT $v_2(T_n)\to v_1(T_n)$ satisfies the condition  $v_1(T_n)/T_n>1$. In the orange area, defined by the condition $V_{\rm 1L}(v_1;T\!=\!0)\geq V_{\rm 1L}(v_2;T\!=\!0)$ with $\lambda_\kappa\leq 0.06$, no EWPT arises because the Universe gets stuck in the phase $v_2$. In the white area $\mu_\kappa^2$ is positive.
In the region on the left of the dashed line, all signal strengths of the Higgs into SM particles are diluted,
for the channel $h\to\kappa\kappa$ is allowed. The strong EWPT found in this area are then ruled out by the present LHC measurements~\cite{Atlas:2015at}. Overall, within the parameter space compatible with DM and the EWSB constraints, the ingredients for electroweak baryogenesis are often realized in the present CHM.

We remark that our result is not a full proof that the model can actually reproduce the measured baryon asymmetry~\cite{Ade:2015xua}. 
By applying straightforwardly the analysis of ref.~\cite{Espinosa:2011eu} to our setup, one would naively reach a positive conclusion. Nevertheless, the expansion velocities~\cite{Moore:1995si,Espinosa:2010hh,Konstandin:2014zta,Megevand:2014dua,Kozaczuk:2015owa} of our EWPT bubbles may not be subsonic (i.e. the bubble speed is smaller than $1/\sqrt{3}$) as assumed in ref.~\cite{Espinosa:2011eu}. Were this the case, the evaluation of the baryon asymmetry would be controversial~\cite{Caprini:2011uz}. This particularly applies to the EWPTs marked as filled circles in fig.~\ref{fig:ewpt}. They  satisfy the runaway condition $\alpha>\alpha_\infty$ with
\begin{eqnarray}
\alpha&\simeq&   \frac{V_{\rm{1L}}(v_2(T_n); T_n)-V_{\rm{1L}}(v_1(T_n); T_n)}{35~ T_n^4} ~,\\
\alpha_\infty&\simeq&    4.9\times 10^{-3} \left(\frac{v_1(T_n)}{T_n}\right)^2~, \label{eq:alphainf}
\end{eqnarray}
which hints at expansion velocities similar to the speed of
light~\cite{Espinosa:2010hh,Caprini:2015zlo}. Consistently, they also
fulfill the microphysical-approach runaway condition $\widetilde
V_{\rm{1L}}(v_1(T_n); T_n) < \widetilde V_{\rm{1L}}(v_2(T_n); T_n)$,
with $\widetilde V$ being the one-loop thermal potential evaluated in
the mean field approximation~\cite{Bodeker:2009qy}~\footnote{We remind
  that these conditions guarantee a runaway behavior only to those
  bubbles that have an initial ultrarelativistic speed. Here we adopt
  the simple criterion $\alpha > \alpha_\infty$, which results
  slightly more stringent than the microphysical one, as noticed also
  in Ref.~\cite{Katz}.  On the other hand, unquestionable
  uncertainties jeopardize these criteria.  Further developments in
  the field are then expected to modify our runaway/non-runaway
  classifications.
}.
Smaller velocities can instead arise for the EWPTs represented by empty circles (satisfying the non-runaway relation $\alpha<\alpha_\infty$), although determining whether such speeds are too large for electroweak baryogenesis would be delicate~\cite{Konstandin:2014zta, Kozaczuk:2015owa}. Due to these uncertainties, we cannot go further than claiming that a quantitative explanation of the observed baryon asymmetry is {\it conceivable} in the parameter points highlighted in fig.~\ref{fig:ewpt}.

\section{Other constraints and future expectations}
\label{sec:constraints}

\noindent We have seen that, for $f\simeq 2.5\div 2.9$ TeV, $m_\kappa \simeq 70\div 120$ GeV, $\lambda_{hk}\simeq 0.2\div 0.4$ and $\lambda_\kappa\simeq 0.01\div 0.06$, which are natural values within our CHM,  the observed DM relic abundance as well the EWPT that is necessary for electroweak baryogenesis are achieved. In this section we check that these ranges of values are not in conflict with present experimental bounds but testable in the forthcoming years.

A common prediction to all CHMs (and models with non-linear realizations of a gauge symmetry in general) is the modification of the Higgs couplings to the SM fermions and gauge bosons. As in the minimal CHM, if we expand eqs.~(\ref{eq:sigma}) and (\ref{eq:yukawa}) to order $\mathcal{O}(v^2/f^2)$,  we obtain the following ratios of the tree level couplings of the Higgs to two SM particles:
\begin{align}
 R_{hVV} &\equiv \frac{g_{hVV}}{g_{hVV}^{SM}} \simeq 1 - \frac{v^2}{2f^2}~,\\
 R_{h\psi\psi} &\equiv \frac{g_{h\psi\psi}}{g_{h\psi\psi}^{SM}} \simeq 1-\frac{3v^2}{2f^2}~,
\end{align}
where $V$ and $\psi$ stand for any electroweak gauge boson and SM fermion, respectively. Even for $f=2.5$ TeV, we obtain $R_{hVV}\sim 0.99$ and $R_{h\psi\psi}\sim 0.98$, therefore well within the current LHC limits~\cite{Atlas:2015at}. Such small deviations from the SM predictions might be however accessible at a future linear collider (see e.g.~ref.~\cite{Klute:2013cx}).
\begin{figure}[t]
  \centering
 \includegraphics[width=\columnwidth]{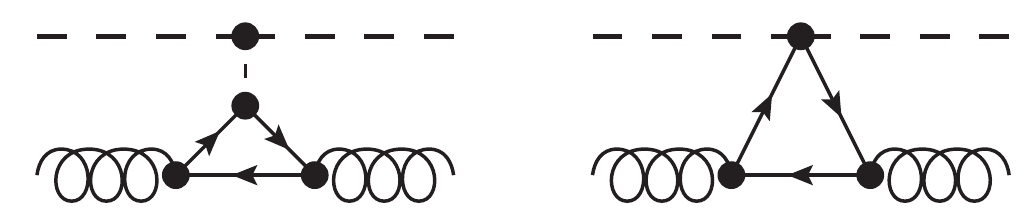}
 \caption{Main diagrams contributing to the scattering of DM particles by nuclei.}\label{fig:dd}
\end{figure}

Concerning direct detection experiments, the two main diagrams contributing to the scattering between DM particles and nuclei are depicted in fig.~\ref{fig:dd}. The corresponding cross section can be parameterized as~\cite{Cline:2013gha}
\begin{equation}\label{eq:xsec}
 \sigma = \lambda_{h}^2\frac{f_N^2}{4\pi}\frac{\mu_r^2 m_n^2}{m_h^4 m_\eta^2}\left[1+\frac{m_\eta^2}{f^2}\right],
\end{equation}
where $m_n$ is the nucleon mass, $\mu_r$ is the reduced mass of the system (with $m_\eta\gg m_n$)
\begin{equation}
 \mu_r= \frac{m_\eta m_n}{m_\eta + m_n}\sim m_n \sim 1\,\text{GeV}~,
\end{equation}
and $f_N\sim 0.3$~\cite{Alarcon:2011zs,Alarcon:2012nr,Kahlhoefer:2015jma}. For the considered ranges of parameter values, eq.~(\ref{eq:xsec}) yields $\sigma\sim 10^{-46}\div10^{-45}$ cm$^2$, depending on the actual value of $f$. These values are around one order of magnitude below the LUX bound in the DM mass range $730\div 960$ GeV~\cite{Akerib:2013tjd}. However, it will be definitely reachable in the new round of data and experiments ~\cite{Cushman:2013zza}~\footnote{This result can be straightforwardly applied to the simpler model $SO(6)/SO(5)$ with $q_L$ transforming in the representation $\mathbf{20}$, which is analogous to our model in the limit $m_\kappa \gg m_\eta$. In this case, the observed DM abundance fixes $m_\eta\simeq 500$ GeV, for which the scattering cross section at zero momentum is larger ($\sim 2 \times10^{-45}$ cm$^2$) but still below the LUX upper limit.}.

\begin{figure*}[t]
  \centering
\hspace{-10mm}\includegraphics[width=\columnwidth]{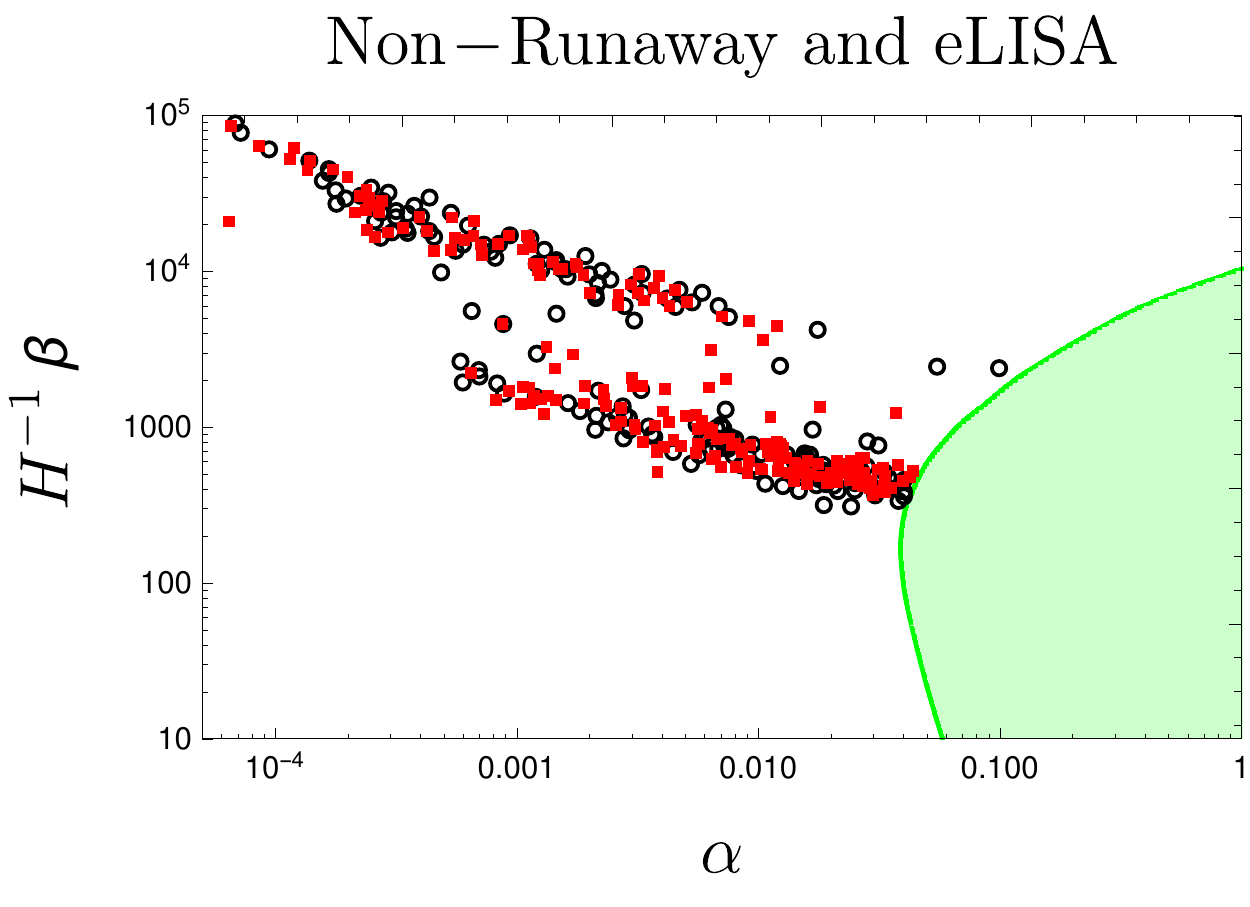}
\includegraphics[width=1.01\columnwidth]{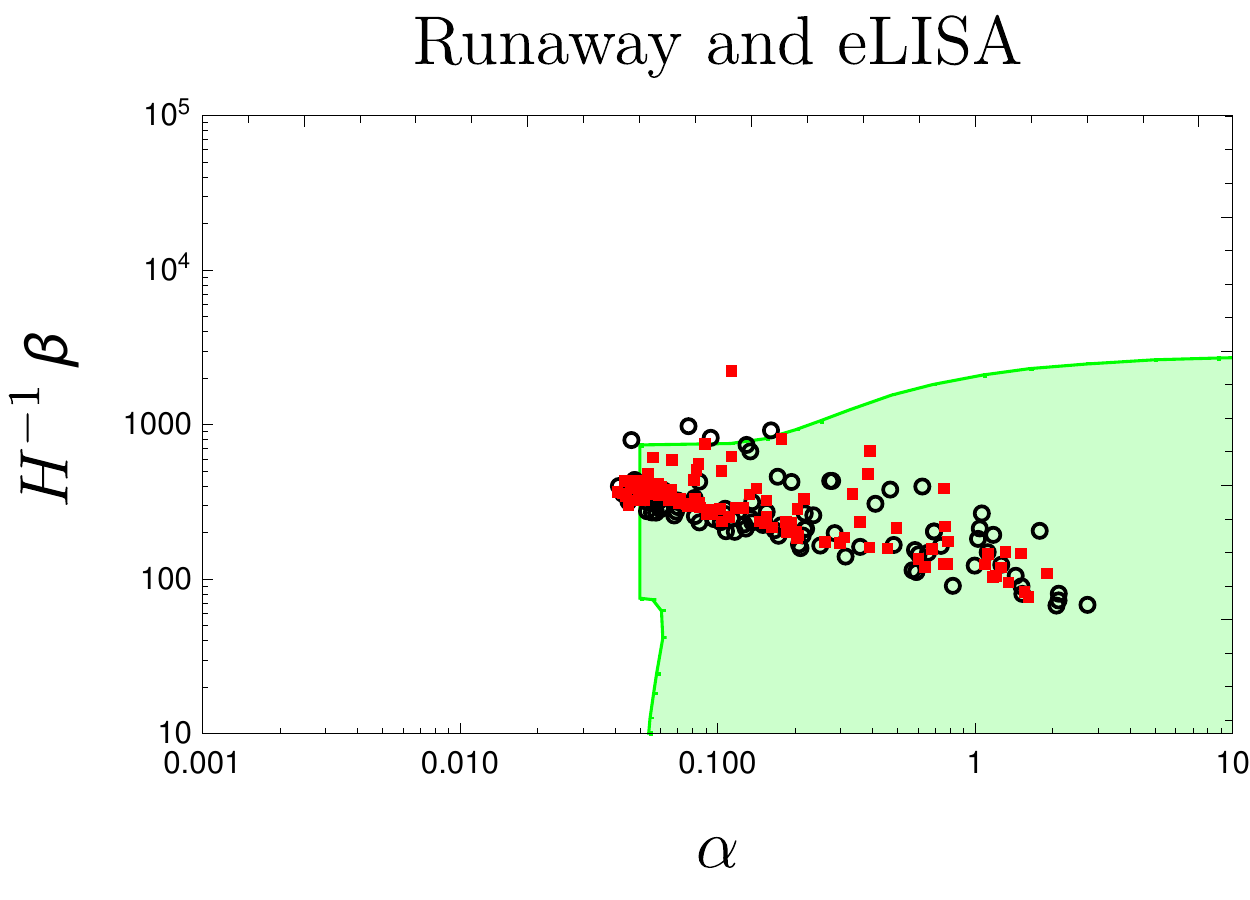}
\caption{The identified first-order EWPTs  with non-runaway (left panel) and runaway (right panel) behavior  in the $\{\alpha,\beta/H\}$ plane.  Black circles and red squares represent the EWPTs of \textit{Regime I} and \textit{Regime II}, respectively. eLISA in the N2A5M5L6 experimental design can test the EWPTs inside the green areas.\label{fig:elisa}}
\end{figure*}
On another front, LHC searches for DM in monojet and $t\bar{t}$ events produced in association with large missing transverse energy do not provide relevant constraints. Indeed, the cross sections of these signals are suppressed by  the small available phase space and the small ratio $v^2/f^2$. On the same vein, searches for double $\eta$ or $\kappa$ production mediated by an off-shell Higgs boson are extremely challenging~\cite{Mambrini:2011ik,Djouadi:2011aa,Djouadi:2012zc,deSimone:2014pda,Craig:2014lda} and hence negligible. Searches for the Higgs boson decaying to non-SM particles~\cite{Aad:2015pla} instead constrain only the small region on the left of the dashed vertical line in fig.~\ref{fig:ewpt}.

A last collider constraint comes from dijet searches. The field $\kappa$ couples linearly to $c$ and $b$ quarks, mainly. Hence, $\kappa$ can be singly produced in hadron colliders, with a subsequent dijet decay. The strongest and updated upper bound on such a process can be found in ref.~\cite{Chala:2015ama} (see also ref.~\cite{Dobrescu:2013coa}). Given the large QCD background, the main constraints at $\mathcal O (100\,$GeV) invariant masses come from UA2~\cite{Alitti:1993pn} and LHC searches for dijet resonances in association with a $W$ or a $Z$ at $\sqrt{s} = 8$ TeV~\cite{TheATLAScollaboration:2013gia}, the latest being dominant for invariant masses below $\lesssim 200$ GeV. In this region, the upper limit on the cross section ranges between $\sim 0.1 \div 0.2$ pb. On another hand, by using \texttt{MadGraph}~\cite{Alwall:2014hca} we find that, even for $\gamma = \zeta = 10$, the production of $\kappa$ in association with a massive gauge boson is smaller than 0.002 pb for any mass in our range of values. The particle $\kappa$ would be then difficult to discover by means of the usual analyses.

A strong EWPT produces a gravitational wave stochastic background whose typical spectrum has a peak at frequencies eLISA is sensitive to~\cite{Seoane:2013qna, webpage}. eLISA might hence be able to probe the CHM under study. We explore this possibility by following ref.~\cite{Caprini:2015zlo}, assuming the plausible ``N2A5M5L6'' eLISA experimental design~\cite{Caprini:2015zlo, APetit}.
Thus, for each point in fig.~\ref{fig:ewpt}, we calculate
$\beta/H\equiv T\partial_{T}(S_3/T)|_{T=T_n}$ (where $S_3$ is the
$O_3$-symmetric bubble action~\cite{Quiros:1994dr}) by running
\texttt{CosmoTransitions}~\footnote{The code determines $T_n$ by
  solving the condition $S(T_n)/T_n=140$~\cite{Wainwright:2011kj}. It
  is then easy to obtain a second temperature, $T_C$, given by
  $S(T_C)/T_C= C$, and use that in the approximation $\beta/H\approx
  T_n (C - 140)/(T_C- T_n)$. We use $C=240$ in our estimates.}. We
display these points in the $\{\alpha,\beta/H\}$ plane, and check
whether they are located inside the pertinent eLISA detection region.
The findings are presented in the left (right) panel of
fig.~\ref{fig:elisa} where the non-runaway (runaway) strong-EWPT
points of {\it Regime I} and {\it Regime II} are displayed as empty
circles and filled squares, respectively. The left (right) panel also
reports the N2A5M5L6 eLISA detection region for non-runaway scenarios
with $T_n\approx 50\,$GeV (for runaway scenarios at $T_n\approx
50\,$GeV and $\alpha_\infty\approx 0.05$) in green color~\footnote{The
  choice of these particular regions is due to the fact that in our
  set of points the nucleation temperature of the EWPT is $30\div
  80\,$GeV, and that the points at the border of detection turn out to
  have $\alpha_\infty\simeq 0.05$. The regions are taken from
  ref.~\cite{Caprini:2015zlo}.}. It results that, among the strong
EWPTs identified in sec.~\ref{sec:ew}, none of those with a
non-runaway behavior is detectable, whereas the runaway ones can be
basically all probed~\footnote{We stress that this sharp conclusion is
  specific to our model where empirically $\alpha_\infty\approx
  \sqrt{\alpha}/2$, whence $\alpha\lesssim 0.05$ for all non-runaway
  cases (defined by the condition $\alpha<\alpha_\infty$). This of
  course depends on the adopted runaway condition and its
  approximation in eq.~\ref{eq:alphainf}. For a much more stringent
  runaway criterion, points of the right panel of fig.~\ref{fig:elisa}
  would certainly move into the detection region of the left
  panel.}. From fig.~\ref{fig:ewpt} we hence deduce that, overall,
eLISA can detect a sizeable fraction of the strong EWPTs predicted
within our model.

\section{Conclusions}
\label{sec:conclusions}

\noindent
We have presented a composite version of the Higgs sector extended with two gauge singlets $\eta$ and $\kappa$ based on the coset $SO(7)/SO(6)$. The embedding of the elementary fermions into appropriate representations of $SO(7)$ can both make $\eta$ stable and generate a negative quadratic term for $\kappa$. As a consequence, this theoretical setup can provide a natural explanation for the observed dark matter abundance and the baryon-antibaryon asymmetry via freeze out and electroweak baryogenesis. We have emphasized that, contrary to its renormalizable counterpart, the dominant terms in the scalar potential are described by only three free parameters, namely the physical mass $m_\eta$ of the dark matter scalar, the physical mass  $m_\kappa$ of the second singlet, and the coupling $\lambda_{h\kappa}$ of the quartic interaction between the Higgs and $\kappa$. We have shown that fulfilling the dark matter relic density observation requires $m_\eta\sim 730\div960$ GeV, irrespectively of $m_\kappa$, which is required to live in the region $50\div 100$ GeV for $\lambda_{h\kappa}\sim 0.15\div0.4$ to trigger the strong first-order electroweak phase transition required by electroweak baryogenesis. We have subsequently studied the implications of this scenario in Higgs physics, dark matter direct detection experiments, LHC searches for dark matter as well as searches for dijet resonances. All together, they are in agreement with current data. The observation (or not) of a dark matter direct detection signal in the sub-TeV region compatible with a total cross section of $10^{-46}\div 10^{-45}$ cm$^2$, possibly followed by the measurement of the gravitational wave stochastic background at eLISA,  will definitely shed light on whether this model is realized in Nature.\\

\vspace{-.5cm}
\section*{Acknowledgements}
We are grateful to Thomas Konstandin and Geraldine Servant for useful comments on the manuscript. IS thanks the hospitality of the DESY Theory Group in the context of the DESY Summer School during the first stage of this work. GN thanks the eLISA cosmology working group for useful discussions and M.~Codiglia for seminal explanations. The work of GN is supported by the Swiss National Science Foundation under grant 200020-155935. The work of IS is supported by RSCF grant 14-22-00161.

\newpage

%

\end{document}